\documentclass[english]{article}
\usepackage[T1]{fontenc}
\usepackage[utf8]{inputenc}
\usepackage{amsthm}
\usepackage{amsmath}
\usepackage{graphicx}
\usepackage{esint}

\makeatletter

\numberwithin{equation}{section}

\usepackage{jheppub}
\usepackage{ae,aecompl}
\renewcommand\[{\begin{equation}}
\renewcommand\]{\end{equation}}
\numberwithin{equation}{section}
\author{Itamar Yaakov}
\affiliation{Department of Physics, \\ Princeton University, Jadwin Hall,\\ Princeton NJ 08544}
\emailAdd{iyaakov@princeton.edu}
\abstract{We give a Seiberg-like dual description of the interacting superconformal infrared fixed point of $\mathcal{N}=4$ gauge theory in three dimensions with vanishing Chern Simons level and $N_c\le N_f<2N_c$ fundamental flavors. These theories are known as "bad" theories due to the existence of unitarity violating monopole operators. We show that, in a dual description, all such operators are realized by free fields and the remainder theory is the Seiberg-like dual previously identified using the type IIB brane construction.}
\keywords{duality, supersymmetry, matrix models}

\makeatother

\usepackage{babel}
\begin{document}

\title{Redeeming Bad Theories}

\maketitle

\section{Introduction}

There has lately been a resurgence of interest in gauge theories with
extended supersymmetry in three dimensions. Much of the focus has
been on superconformal theories in which there is a Chern-Simons kinetic
term for the gauge field, instead of the usual Yang-Mills term. The
search for such theories was initiated in \cite{Schwarz:2004yj} and
has yielded an array of superconformal actions with $\mathcal{N}=3,4,5,6,8$
supersymmetry in the three dimensional sense. The most thoroughly
studied example is ABJM theory \cite{Aharony:2008ug}, which, for
special values of the Chern-Simons level ($k$), has maximal supersymmetry.
Some of these theories, and ones without a Chern-Simons term, can
be constructed as low energy effective theories on a stack of D-branes
in a Hanany-Witten type setup in type IIB string theory (\cite{Hanany:1996ie})
or on M2 branes in M theory. Theories with $k=0$ are inevitably strongly
coupled in the IR because $g_{YM}^{2}$ has mass dimension $1$, but
some are believed to flow to interacting superconformal fixed points
\cite{Intriligator:1996ex}. 

Much has been learned about the low energy dynamics of 3d gauge theories
with $\mathcal{N}\ge2$ supersymmetry by the combined use of holomorphy
and localization. The most exciting discovery has been a web of dualities
which relate interacting superconformal IR fixed points of different
UV theories. Examples include mirror symmetry of $\mathcal{N}=4$
theories at $k=0$ \cite{Intriligator:1996ex}, and its extensions
to $\mathcal{N}=2$ and to some theories with $k=\pm1$ \cite{Jensen:2009xh,deBoer:1996ck,deBoer:1996mp,deBoer:1997ka,Aharony:1997bx}.
This was shown to descend, in some cases, from s-duality of the type
IIB construction \cite{Hanany:1996ie}. A somewhat different class
of examples are known collectively as Seiberg-like duality because
the rank of the dual gauge group depends on the number of flavors
in the original theory as in 4d Seiberg duality \cite{Seiberg:1994pq}.
These include Giveon-Kutasov duality \cite{Giveon:2008zn} ($\mathcal{N}=2,3$
and $k\ne0$) and Aharony duality \cite{Aharony:1997gp} ($\mathcal{N}=2$
and $k=0$). An $\mathcal{N}=4$ Seiberg-like duality can also be
argued for by considering the effect of a set of brane moves as described
in \cite{Hanany:1996ie}. It is known, however, that the dualities
implied by reading off the gauge theory associated to the initial
and final brane configurations are wrong. The dynamics of the low
energy degrees of freedom on the Coulomb branch is incorrectly accounted
for. The correct behavior in the $\mathcal{N}=2$ case is given by
the Aharony dual which includes extra dual fields and a somewhat complicated
superpotential (the theories involved in Giveon-Kutasov duality do
not have Coulomb branches and can be read off from the branes). 

We will present a proposal for the case of $\mathcal{N}=4$ and $k=0$.
These theories have a large moduli space of supersymmetric vacua where
the gauge group is partially Higgsed. We will give a dual description
of the interacting fixed point at the origin of the Higgs branch.
Specifically, we will propose that the $U(N_{c})$ theory with $N_{c}\le N_{f}<2N_{c}$
is dual to the $U(N_{f}-N_{c})$ theory with $N_{f}$ flavors and
$2N_{c}-N_{f}$ additional free (twisted) hypermultiplets. The usual
meson fields of Seiberg duality are absent. The larger amount of supersymmetry
present in these examples makes the analysis easier, but also implies
a larger coulomb branch \cite{Seiberg:1996nz}. The interacting part
of the UV actions is the one that can be read off from the type IIB
branes \cite{Hanany:1996ie}. The decoupled fields can be argued for
by considering the light fields at various point on the moduli space
as was done in \cite{Gaiotto:2008ak} for $N_{f}=2N_{c}-1$. The analysis
is very similar to the one presented in 4d in \cite{Argyres:1996eh}. 

In an IR phase with unbroken gauge symmetry, 3d gauge theories admit
pointlike defects known as monopole operators \cite{Borokhov:2002ib}.
In a theory with $\mathcal{N}\ge2$ supersymmetry and $k=0$ these
can be promoted to chiral operators whose quantum numbers, including
the R-charge, can be systematically deduced from the UV theory \cite{Borokhov:2002cg,Bashkirov:2010kz}.
The superconformal algebra implies that the dimension of these operators,
or any chiral operator, in the IR theory is equal to their charge
under a particular R-symmetry whose current sits in the same multiplet
as the energy momentum tensor. This distinguished R-symmetry may differ
from the UV R-symmetry, for example by mixing with a flavor symmetry.
The IR R-charge can still, in some cases, be recovered by using $Z$
minimization \cite{Jafferis:2010un}. It is also possible that the
distinguished R-symmetry is not part of the UV symmetry algebra (it
may be accidental). For example, the R-charge implied by the UV algebra
may be in conflict with unitarity. In this situation it is not in
general known how to recover the correct R-charge. 

Quiver theories with $\mathcal{N}=4$ supersymmetry and $k=0$ can
be classified according to their spectrum of monopole operators \cite{Gaiotto:2008ak}.
Quivers with a ``standard'' fixed point, where all monopole operators
have dimension $\ge1$, are called ``good'', and those with monopole
operators of dimensions $1/2$, which implies that they are realized
by free fields, (but none with vanishing or negative dimension) are
called ``ugly''. Theories with monopole operators of vanishing or
negative dimensions, which implies unitarity violation, are called
``bad''. The distinguished IR R-symmetry for a ``bad'' theory
is not visible in the UV \cite{Gaiotto:2008ak}. It can be shown that
convergence of the $S^{3}$ partition function, computed using localization,
is correlated with this classification such that the partition function
for ``bad'' theories is divergent \cite{Kapustin:2010mh}. This
can be attributed to the fact that the correct evaluation of the partition
function using localization is predicated on knowing the correct R-charge.
This data is used to write down the action of the theory on $S^{3}$
\cite{Festuccia:2011ws}. We will see that the partition function
for a ``bad'' theory can still be regularized and compared to its
dual. 

In Section \ref{sec:Proposal-and-preliminary}, we describe a proposal
for the Seiberg-like duality of $\mathcal{N}=4$ theories. The duality
relates ``bad'' theories to ``good'' ones. We compare the global
symmetries and moduli space of the dual pairs. In Section \ref{sec:The--partition}
we identify a regularized version of the squashed sphere partition
function of a ``bad'' theory (this includes the round sphere as
a special case). We show that the dual pairs have the same partition
function for all values of the available deformation parameters. We
end with a short discussion of possible further checks of the duality
and related questions.

\section{\label{sec:Proposal-and-preliminary}Proposal and preliminary checks}

\begin{figure}
\begin{centering}
\includegraphics[width=15cm]{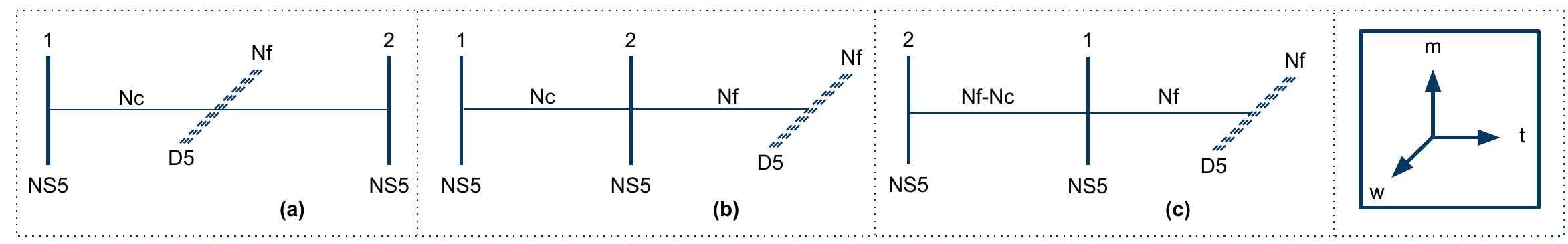} 
\par\end{centering}

\caption{Brane manipulations in type IIB string theory which yield a possible
Seiberg-like dual \cite{Hanany:1996ie}. Solid vertical lines are
NS5 branes. Horizontal lines are coincident D3 branes. Dashed lines
are D5 branes. The legend indicates the compactification direction
(t or $x_{6}$) and the directions of possible triplet mass (m) terms
(3,4,5), and possible triplet FI (w) terms (7 8 9). Directions (0
1 2) are common to the world volume of all branes and are suppressed.
We first move $N_{f}$ D5 branes through the right NS5 brane, creating
$N_{f}$ D3 branes in the process. We then exchange the two NS5 branes,
changing the number of suspended D3 branes in the interval.\label{fig:Naive_duality}}
\end{figure}

We will attempt to partially characterize the IR fixed point at the
origin of the Higgs branch of $\mathcal{N}=4$ $U(N_{c})$ 3d gauge
theories with $N_{c}\le N_{f}<2N_{c}$ fundamental flavors and no
Chern-Simons term. In the classification of \cite{Gaiotto:2008ak},
such theories are called ``bad''. An $\mathcal{N}=4$ theory, in
the three dimensional sense, has $8$ real supercharges. A fundamental
flavor includes two $\mathcal{N}=2$ chiral superfields transforming
in the fundamental and anti-fundamental representations respectively.
The action for the $\mathcal{N}=4$ theories, including the real mass
and Fayet-Iliopoulos deformations, can be found in \cite{Kapustin:2010xq}.
The construction in terms of D-branes in type IIB string theory was
described in \cite{Hanany:1996ie}. This work also described a set
of brane moves which imply a Seiberg-like duality for these theories.
Figure \ref{fig:Naive_duality} illustrates the relevant configuration.
The conclusion in \cite{Hanany:1996ie} was that the convergence of
two $NS5$ branes in spacetime, as require by the brane moves, may
create too severe a singularity which would then invalidate the duality.

\subsection{A Seiberg-like duality}

The simplest example of a ``bad'' theory, which is, however, not
part of our proposal, is $U(1)$ gauge theory with $\mathcal{N}=4$
supersymmetry and no matter. Explicit analysis, which is possible
because the theory is free, shows that the low energy action is that
of a free twisted hypermultiplet constructed by dualizing the gauge
field $A_{\mu}$ into an additional scalar \cite{Gaiotto:2008ak}.
The R-symmetry acting on this hypermultiplet is not visible in the
UV action (hence the fixed point is not ``standard''). The UV action
admits monopole operators which have IR conformal dimension $0$ from
the point of view of the UV algebra. One can try and identify the
free hypermultiplet with these operators, but the duality obscures
the symmetry action.

An ``ugly'' theory is one with $N_{f}=2N_{c}-1$ fundamental flavors.
There are no unitarity violating monopole operators in these theories,
but there are operators of R-charge $1/2$. These parametrize a decoupled
free sector of the low energy theory, consisting of a single free
hypermultiplet. The remainder has a dual description in terms of a
$U(N_{c}-1)$ theory with $2N_{c}-1$ massless hypermultiplets. The
remainder theory is ``good'' and the duality is reminiscent of Seiberg
duality in that the dual theory has a gauge group $U(N_{f}-N_{c})$
\cite{Gaiotto:2008ak}.

A similar attempt to describe an interacting ``bad'' theory fails
because the theory is neither free nor has any easily identifiable
free sectors although it is expected, on general grounds, that any
operator which dips below the unitarity bound is actually free \cite{Seiberg:1994pq}.
A general feature of these theories is that the gauge group cannot
be completely Higgsed by giving vevs to the matter multiplet scalars
(compatible with the scalar potential). As a result, even vacua far
away on the Higgs branch contain massless vector multiplets. There
could still be a singularity, and hence an interacting fixed point,
at the origin of this branch, but it would be reasonable to expect
that there is also a decoupled free sector.

The resolution we propose is as follows
\begin{itemize}
\item Theories with $N_{f}=N_{c}$ are expected to have a smooth moduli
space. The IR free theory on this space can be equivalently described
using $N_{f}$ twisted hypermultiplets corresponding to monopole operators. 
\item Theories with $N_{c}<N_{f}<2N_{c}$ have an IR fixed point which includes
an interacting sector and a decoupled free sector. The decoupled sector
can be described by $2N_{c}-N_{f}$ free hypermultiplets. The interacting
sector has a Seiberg-like dual description as the IR fixed point at
the origin of the Higgs branch of the $\mathcal{N}=4$ $U(N_{f}-N_{c})$
theory with $N_{f}$ fundamental hypermultiplets. The Seiberg-like
dual is ``good'' and hence has no unitarity violating monopole operators. 
\end{itemize}
We will provide some standard evidence for this proposal: matching
of the global symmetries, the dimension of the moduli space (actually
the Higgs branch metric) and of the regularized $S_{b}^{3}$ (i.e.
the squashed sphere) partition function. Note that the case $N_{f}=2N_{c}-1$
is well established. The equality of the $S^{3}$ partition function,
which is convergent in this case, was shown in \cite{Kapustin:2010mh}.
Theories with $N_{f}<N_{c}$ are also ``bad'', but the duality proposal
does not apply to them (the dual gauge group would have negative rank).
These theories are not expected to have interacting fixed points.

\subsection{Global symmetries }

In three dimensions, an $\mathcal{N}=4$ $U(N_{c})$ gauge theory,
with $N_{f}$ massless fundamental hypermultiplets and vanishing Chern-Simons
level has a global bosonic symmetry group 
\[
SU(N_{f})_{\text{flavor}}\times SO(4)_{\text{R}}\times U(1)_{J}
\]
The first factor is a flavor symmetry which rotates the fundamental
and anti-fundamental chiral superfields in opposite directions. The
second factor is an R-symmetry under which the supercharges transform
as a four-vector. $U(1)_{J}$ is a topological symmetry whose current
is $\star tr(F)$, which is divergenceless due to the Bianchi identity.
Fundamental fields are not charged under $U(1)_{J}$, but monopole
operators may be. 

When written in $\mathcal{N}=2$ language, only a $U(1)_{R}\times U(1)_{A}$
subset of the R-symmetry group is visible. The $U(1)_{R}$ part is
an R-symmetry of the $\mathcal{N}=2$ algebra. In a CFT with $\mathcal{N}=2$
supersymmetry, a chiral superfield satisfies \cite{Aharony:1997bx}
\[
D=|R|\ge\frac{1}{2}
\]
where $D$ is the conformal dimension of the lowest component of the
superfield and $ $$R$ is its charge under the R-symmetry that sits
in the same superconformal multiplet as the stress-energy tensor.
This may or may not be identified with the charge under the $U(1)_{R}$
for the UV action. A chiral operator with $D=1/2$ is realized by
a free field. A gauge invariant chiral operator with $D<1/2$ violates
the unitarity bound. 

Our Seiberg-like duals trivially have the same global symmetry algebra
except for the ``accidental'' currents coming from the decoupled
sector. We do not expect to be able to see these currents in the original
action.

\subsection{Moduli space}

The classical moduli space of the original theory is a mixed Coulomb-Higgs
branch where vevs for the scalars in the matter hypermultiplets partially
Higgs the gauge group. At a generic point on this space, the $N_{c}\times N_{f}$
dimensional matrix of scalar vevs for the fundamental chirals has
rank $r=N_{f}/2<N_{c}$ for $N_{f}$ even and $r=(N_{f}-1)/2$ for
$N_{f}$ odd (the matrix of anti-fundamentals has the same rank \cite{Gaiotto:2008ak}).
The rank of the unbroken gauge group is $N_{c}-r=N_{c}-N_{f}/2$ for
$N_{f}$ even and $N_{c}-r=N_{c}-(N_{f}-1)/2$ for $N_{f}$ odd. The
space of scalar vevs (the Higgs branch) is a hyper-Kähler manifold
of dimension (we count quaternionic dimensions throughout) $N_{c}(N_{f}-N_{c})$
given by the hyper-Kähler reduction of the (flat) space of hypermultiplet
scalars by the gauge group $G$. The low energy action for the hypermultiplets
is a hyper-Kähler sigma model \cite{1987CMaPh.108..535H}. 

There is also a Coulomb branch emanating from this space. The dimension
of this branch (for $N_{f}$ even) is $(N_{c}-N_{f}/2)\le f\le N_{c}$
where the lower bound is the generic dimension on the bulk of the
space and the upper bound occurs at the origin, where all the hypermultiplet
scalars vanish. Out on this branch, the gauge group is generically
further (spontaneously) broken to $U(1)^{f}$. Far away on the Coulomb
branch, the light vector multiplets can be dualized into additional
hypermultiplets. The resulting low energy theory is again a hyper-Kähler
sigma model. 

In the quantum theory the space above may receive corrections. It
is not possible for a superpotential to be generated which would lift
a part of the space, but there can be corrections to the metric \cite{Seiberg:1996nz}.
$\mathcal{N}=4$ supersymmetry guarantees that the moduli space is
a hyper-Kähler manifold after all quantum corrections are taken into
account. Our duality conjecture implies that the moduli space for
theories with $N_{c}<N_{f}<2N_{c}$ has a $2N_{c}-N_{f}$ dimensional
component described by the decoupled hypermultiplets and a remainder:
an $N_{f}-N_{c}$ dimensional Coulomb branch and an $(N_{f}-N_{c})(N_{f}-(N_{f}-N_{c}))=N_{c}(N_{f}-N_{c})$
dimensional Higgs branch which meet at a singularity. 

The equivalence of the Higgs branches for the original and dual theories,
including the metric, is known from the analysis of 4d $\mathcal{N}=2$
theories \cite{Antoniadis:1996ra}. In fact, the Higgs branch has
been shown to coincide with the cotangent bundle of the complex grassmanian
$G_{N_{c},N_{f}}$ and the equivalence of the Higgs branch metrics
is an extension of grassmanian duality 
\[
G_{N_{c},N_{f}}=G_{N_{f}-N_{c},N_{f}}
\]
The fact that the theory does not have vacua where the gauge group
is completely Higgsed provided the motivation for the Seiberg-like
duality of the ``ugly'' theory in \cite{Gaiotto:2008ak}. When $N_{c}<N_{f}\le2N_{c}-2$,
the unbroken gauge symmetry at a generic point on the mixed Higgs/Coulomb
branch is $U(1)^{N_{c}-N_{f}/2}$ (for $N_{f}$ even). Comparing with
the ``ugly'' case, where the unbroken gauge group is $U(1)$, we
might be led to believe that the light $U(1)$ gauge multiplets can
be dualized into $N_{c}-N_{f}/2$ hypermultiplets. However, the points
where all gauge multiplet moduli vanish have enhanced (non-abelian)
gauge symmetry, and strong gauge dynamics, which may drastically alter
the action for these hypermultiplets. From the study of $\mathcal{N}=2$
theories, it is known what such dynamics may require the introduction
of additional fields even when the unbroken gauge symmetry is abelian
\cite{Aharony:1997bx} (these are the $V_{\pm}$ fields of \cite{Aharony:1997gp}).
It seems plausible that a similar effect requires the use of $2N_{c}-N_{f}$
hypermultiplets for $\mathcal{N}=4$. This has the advantage of making
the dimension of the branch emanating from the origin of the Higgs
branch the same for a theory and its Seiberg-like dual.

\subsection{\label{sub:Flowing-from-ugly}Flowing from ``ugly'' to ``bad''}

A ``bad'' theory can be reached by an RG flow from a ``good''
or ``ugly'' theory by giving large masses to some of the hypermultiplets.
Integrating out the massive matter does not induce a Chern-Simons
term for the gauge field as long as the masses for the two Dirac fermions
in a hypermultiplet have opposite signs, which is the case for the
real mass deformations of $\mathcal{N}=4$ \cite{Aharony:1997bx}.
Starting from the Seiberg-like duality of the ``ugly'' theory and
giving a large mass to a single hypermultiplet on both sides we end
up with a $U(N_{c})$ theory with $2N_{c}-2$ flavors on one side
and a $U(N_{c}-1)$ theory with $2N_{c}-2$ flavors and a single free
hypermultiplet on the other. These two theories are not dual as we
will see. Instead, there is a subtlety related to following the correct
vacuum across the duality. It turns out that in the vacuum dual to
the origin of the $U(N_{c})$ theory the dual gauge group is partially
Higgsed to $U(N_{c}-2)\times U(1)$ (with $2N_{c}-2$ massless flavors
charged under the first factor and a single massless hypermultiplet
charged under the second)%
\footnote{We would like to thank Ofer Aharony for suggesting this possibility.%
}. This comes about by adding a suitable vev for one of the vector
multiplet moduli. The $U(1)$ factor (and the charged hypermultiplet)
is then dualized into a twisted hypermultiplet. This duality is actually
mirror symmetry, and is described in detail in \cite{Kapustin:1999ha}.
The resulting dual theories then coincide with our proposal. In the
following section, we motivate this property of the dual vacuum from
the partition function of the theory on the squashed three sphere.

\section{\label{sec:The--partition}The $S_{b}^{3}$ partition function}

An $\mathcal{N}=2$ theory with a conserved $U(1)_{R}$ symmetry can
be coupled to a supergravity multiplet using the R-multiplet \cite{Dumitrescu:2011iu,Festuccia:2011ws,Komargodski:2010rb}.
It is possible to then put the theory on a squashed three sphere while
preserving some of the fermionic symmetries \cite{Hama:2011ea,Imamura:2011wg}.
The partition function on this compact space ($Z_{b}$) is finite
and unambiguous, after a suitable subtraction, and can be calculated
using a localization procedure similar to the ones used in \cite{Kapustin:2009kz,Pestun:2007rz}.
Specializing to the $\mathcal{N}=4$ case with gauge group $U(N_{c})$
and $N_{f}$ fundamental flavors, $Z_{b}$ is a function of the real
mass deformations $m_{a}$ and the Fayet-Iliopoulos (FI) parameter
$\eta$. The parameters $m_{a}$ can be viewed as the constant (on
the squashed sphere) value of the scalar $\sigma$ in a background
vector multiplet introduced by weakly gauging the $SU(N_{f})$ flavor
symmetry (note that the $m_{a}$ sum to $0$). The FI parameter can
be similarly introduced by weakly gauging the $U(1)_{J}$ symmetry.
The partition function, deformed by the $m_{a}$ and $\eta$, can
be computed using a matrix model derived from the localization procedure.
A necessary condition for the proposed duality to hold is that $Z_{b}$
should agree, as a function of the deformations, for the dual theories
(see \cite{Kapustin:2010mh,Kapustin:2010xq}). In this section, we
test this agreement for the Seiberg-like duality introduced above. 

The partition function on the round three sphere for the class of
theories under consideration formally diverges. This is directly correlated
with the unitarity violating dimensions of monopole operators \cite{Kapustin:2010xq}.
We will see that it can still be defined by analytic continuation
in certain deformation parameters. These are the would be anomalous
dimensions (or corrected IR R-charges) of the various chiral multiplets
\cite{Jafferis:2010un,Willett:2011gp}. It should be noted that the
analytic continuation presented in this section is not associated
with a physical correction to the R-charges; It is merely a mathematical
trick used to regulate the partition function. This is sufficient
in order to match the dual partition functions as meromorphic functions
of the deformation parameters \cite{Willett:2011gp}. As a consequence
of this, some of the R-charge assignments made here will not coincide
with the physical dimension of the operators. Nevertheless, the partition
function with the deformations corresponding to the visible UV symmetry
currents should still match.

\subsection{The matrix model}

Localization can be used to reduce the path integral calculation of
the partition function on the squashed sphere to a matrix model. The
derivation for the round sphere can be found in \cite{Kapustin:2009kz}
(see \cite{Marino:2011nm} for a nice review) and for the squashed
sphere in \cite{Hama:2011ea,Imamura:2011wg}. It can be shown that
the matrix integral corresponding to a ``bad'' $\mathcal{N}=4$
theory diverges \cite{Kapustin:2010xq}. The physical interpretation
of this fact is that the coupling to the round sphere, achieved using
the UV R-multiplet, does not correctly capture the R-charges (equivalently,
conformal dimensions) of all operators at the IR fixed point. The
existence of monopole operators of vanishing or negative R-charge
(from the UV point of view) then causes the divergence. The resulting
matrix model and integral are seemingly meaningless. However, it is
known from the study of $\mathcal{N}=2$ theories that coupling the
theory using the wrong R-charge merely sets the imaginary part of
some deformation parameters to unphysical values (from the point of
view of the flat space conformal field theory) \cite{Jafferis:2010un}.
The mixing of the flavor symmetries with the UV R-symmetry gives an
imaginary contribution to the mass parameters $\mu_{a}$ and $\nu_{a}$
(introduced below) and it has been argued that mixing with the $U(1)_{J}$
symmetry induces a similar contribution to $\eta$. The correct value
of the partition function can be recovered by using a ``trial''
R-charge and extremizing the resulting integral \cite{Jafferis:2010un}.
The use of trial R-charges is possible when the superconformal R-symmetry
gets contributions from abelian flavor symmetries or the $U(1)_{J}$
current. There is no known way of incorporating the change to the
R-charges of monopoles operators or from accidental symmetries. 

Luckily, finding the correct R-charge is not necessary if one only
wants to compare the partition function of dual theories. The comparison
depends only on the correct identification of the UV symmetries and
on being able to evaluate $Z_{b}$ in some, not necessarily physical
(in the sense of flat space), region of the parameter space. One expects
that the analytic properties of $Z_{b}$ ensure that its value will
match at the physical points. By physical points, we mean ones where
all fields are coupled to the supergravity multiplet using their IR
R-charges. The $S^{3}$ partition function at the physical points
has an interpretation in terms of the entanglement entropy across
a circle in the flat space CFT \cite{Casini:2011kv}. We will assume
that the ``incorrect'' R-charge assignment used to put a ``bad''
$\mathcal{N}=4$ theory on the squashed sphere can be undone in this
manner. We have assumed, as part of the proposal, that the dimensions
of some of the monopole operators in the original theory are exactly
$1/2$. We will not be able to impose this restriction in the expression
for the partition function of the original theory, but doing so in
the expression for the dual is trivial (the hypermultiplets describing
the monopoles are decoupled from the rest of the matrix model). 

We will work with a (matrix) integral which generalizes the calculation
on the squashed sphere. This is the class of hyperbolic gamma function
integrals introduced in \cite{VanDeBult2008}. As explained in \cite{Willett:2011gp},
the matrix model for an $\mathcal{N}\ge2$ $U(N_{c})$ gauge theory
with $N_{c}\ge N_{f}$ can be written in terms of the following integral
(the notation comes from \cite{VanDeBult2008}) 
\[
I_{n,(2,2)}^{m}(\mu;\nu;\lambda)=\frac{1}{\sqrt{-\omega_{1}\omega_{2}}^{n}n!}\int_{C^{n}}\prod_{1\leq j<k\leq n}\frac{1}{\Gamma_{h}(\pm(x_{j}-x_{k}))}\prod_{j=1}^{n}\bigg(e^{\frac{\pi i\lambda x_{j}}{\omega_{1}\omega_{2}}}\prod_{a=1}^{n+m}\Gamma_{h}(\mu_{a}-x_{j})\Gamma_{h}(\nu_{a}+x_{j})dx_{j}\bigg)
\]
The function $\Gamma_{h}(z|\omega_{1,}\omega_{2})$ is called a hyperbolic
gamma function and satisfies (the dependence on $\omega_{1,2}$ will
often be suppressed)
\begin{align}
\Gamma_{h}(z+\omega_{1})=2\sin(\frac{\pi z}{\omega_{2}})\Gamma_{h}(z)\notag\label{hgam}\\
\Gamma_{h}(z+\omega_{2})=2\sin(\frac{\pi z}{\omega_{1}})\Gamma_{h}(z)\\
\Gamma_{h}(z)\Gamma_{h}(\omega_{1}+\omega_{2}-z)=1\notag
\end{align}
An explanation of the relationship to the squashed sphere partition
function and some identities for the hyperbolic gamma function can
be found in \cite{Benini:2011mf}. The parameters for an $\mathcal{N}=4$
theory are identified as \cite{Willett:2011gp} 
\[
n=N_{c},\;\;\; m=N_{f}-N_{c},\;\;\;\mu_{a}=\frac{\omega}{2}-m_{a},\;\;\;\nu_{a}=\frac{\omega}{2}+m_{a}\;\;\;\lambda=-2\eta
\]
where 
\[
\omega=\frac{\omega_{1}+\omega_{2}}{2}
\]
Note that the factors of $\omega/2$ represent the canonical dimension
of the chiral multiplets (i.e. $1/2$). The elements of the integral
correspond to the path integral for the theory as follows
\begin{itemize}
\item The integration is over the $N_{c}$ ``Coulomb branch'' moduli (on
the sphere) identified with the Cartan elements of the constant component
of $\sigma$, where $\sigma$ is a scalar in the dynamical vector
multiplet. 
\item $\prod_{1\leq j<k\leq n}\frac{1}{\Gamma_{h}(\pm(x_{j}-x_{k}))}$ represents
the determinant coming from fluctuations of the fields in the dynamical
vector multiplet.
\item $\prod_{a=1}^{n+m}\Gamma_{h}(\mu_{a}-x_{j})\Gamma_{h}(\nu_{a}+x_{j})$
is the matter determinant, including the dependence on real mass parameters,
of which there are two sets, $\mu_{a},\nu_{a}$, for an $\mathcal{N}=2$
theory with no superpotential and one set, $m_{a}$ for the $\mathcal{N}=4$
theory. 
\item $e^{\frac{\pi i\lambda x_{j}}{\omega_{1}\omega_{2}}}$ is the contribution
of an FI term. 
\end{itemize}
The parameters $\omega_{1,2}$ take the values $(i,i)$ for the round
sphere, and $(ib,i/b)$ for the squashed sphere. The positive number
$b$ is the squashing parameter \cite{Hama:2011ea}. The integral
$I_{n,(2,2)}^{m}(\mu;\nu;\lambda)$ can be extended to a meromorphic
function of the deformation parameters \cite{VanDeBult2008}. Moreover,
the points where $ $$\Im(\mu_{a})=\Im(\nu_{a})=\omega/2$ and $\Re(\nu_{a})=-\Re(\mu_{a})=m_{a}$
are regular%
\footnote{\label{fn:Regularity}In the language of \cite{VanDeBult2008}, the
parameters must satisfy (note $\tau=\omega$): 
\[
\left(\mu,\nu\right)\in\mathcal{B}_{N_{f},N_{f}}^{\omega},\qquad\left(\omega-\mu,\omega-\nu\right)\in\mathcal{B}_{N_{f},N_{f}}^{\omega}
\]
 and 
\[
\left(\mu,\nu,\pm2\eta\right)\in\mathcal{D}_{N_{c},(N_{f},N_{f})}^{\omega},\qquad\left(\omega-\mu,\omega-\nu,\pm2\eta\right)\in\mathcal{D}_{N_{f}-N_{c},(N_{f},N_{f})}^{\omega}
\]
where we have included the conditions for having a well defined integral
for the partition function of the Seiberg-like dual introduced below.
For an arbitrary positive squashing parameter $b$, the phase convention
being used is such that the phase is zero on the positive real line
and the branch cut is along the negative imaginary axis ($\phi_{+}=\phi_{-}=\pi/2)$.
The first set of conditions are satisfied identically for any choice
of $m_{a}$ and $\eta$. The second set is equivalent to $\beta_{\pm}^{(1)}=\arg\left(\left(N_{f}-2N_{c}+2\right)\omega\pm2\eta\right)\ne-\pi/2$
and $\beta_{\pm}^{(2)}=\arg\left(\left(2N_{c}+2-N_{f}\right)\omega\pm2\eta\right)\ne-\pi/2$.
Hence the only problem that could arise is when $\eta=0$ and $\beta^{(1,2)}$
can be $-\pi/2$ or undefined. Comparison with the case of the abelian
theory with no flavors suggests that there may be a delta function
at $\eta=0$. This seems likely when $N_{f}=2N_{c}-2$ and the asymptotic
behavior is such that the integrand goes to a constant at infinity.
However, when $N_{f}$ is even, the partition functions we find will
have an ordinary pole when continued to $\eta=0$. %
}. We will \emph{define} the squashed sphere partition function of
a ``bad'' $\mathcal{N}=4$ theory with $N_{f}\ge N_{c}$ using the
value of $I_{n,(2,2)}^{m}(\mu;\nu;\lambda)$
\[
Z_{N_{c},N_{f}}^{\mathcal{N}=4}\left(\{m_{a}\};\eta\right):=I_{N_{c},(2,2)}^{N_{f}-N_{c}}(\frac{\omega}{2}-m_{a};\frac{\omega}{2}+m_{a};-2\eta)
\]
For a single free hypermultiplet with R-charge $1/2$ and a real mass
parameter $m$ 
\[
Z_{\text{hyper}}^{\mathcal{N}=4}\left(m\right)=\Gamma_{h}\left(\frac{\omega}{2}\pm m\right)
\]

Note that we have implicitly used the freedom to change the imaginary
parts of $\mu_{a}$ and $\nu_{a}$ to achieve convergence of the integral
and defined the $\mathcal{N}=4$ partition function by analytic continuation.
The value of the partition function on the round sphere matches the
one given in \cite{Kapustin:2010mh}. A more general class of integrals
can be used to represent the $N_{f}<N_{c}$ partition function. These
are the $J{}_{n,(s_{1},s_{2}),t}$ type integrals of \cite{VanDeBult2008}.
However, for the physical values $s_{1}=s_{2}=N_{f}<n=N_{c}$ and
$t=0$ the partition function thus defined vanishes identically (away
from $\eta=0$).

\subsection{Seiberg-like duality}

The integrals introduced above satisfy the following identity as meromorphic
functions of the deformation parameters \cite{VanDeBult2008}

\[
I_{n,(2,2)}^{m}(\mu;\nu;\lambda)=I_{m,(2,2)}^{n}(\omega-\nu;\omega-\mu;-\lambda)\prod_{a,b=1}^{n+m}\Gamma_{h}(\mu_{a}+\nu_{b})\times\notag
\]

\begin{equation}
\times\Gamma_{h}\left((m+1)\omega-\frac{1}{2}\sum_{a=1}^{n+m}(\mu_{a}+\nu_{a})\pm\frac{1}{2}\lambda\right)c\left(\lambda\sum_{a=1}^{n+m}(\mu_{a}-\nu_{a})\right)\label{eq:duality_relation}
\end{equation}
where 
\[
c(x)=\exp\left(\frac{i\pi x}{2\omega_{1}\omega_{2}}\right)
\]

This identity represents the equality of partition functions in a
Seiberg-like duality \cite{Willett:2011gp} (note that $I_{m,(2,2)}^{n}$
is $1$ for $m=0$). For ``ugly'' theories, the duality takes the
form (we write $Z_{N_{c},N_{f}}$)
\begin{equation}
Z_{N_{c},2N_{c}-1}^{\mathcal{N}=4}\left(\{m_{a}\};\eta\right)=Z_{N_{c}-1,2N_{c}-1}^{\mathcal{N}=4}\left(\{m_{a}\};-\eta\right)Z_{\text{hyper}}^{\mathcal{N}=4}\left(\eta\right)Z_{\text{background FI}}\left(\{m_{a}\};\eta\right)\label{eq:N4_Ugly_Partition}
\end{equation}
which descends from the identity by way of

\begin{equation}
I_{N_{c},(2,2)}^{N_{c}-1}\left(\frac{\omega}{2}-m_{a};\frac{\omega}{2}+m_{a};-2\eta\right)=I_{N_{c}-1,(2,2)}^{N_{c}}\left(\frac{\omega}{2}-m_{a};\frac{\omega}{2}+m_{a};2\eta\right)\Gamma_{h}\left(\frac{\omega}{2}\pm\eta\right)c\left(4\eta\sum_{a=1}^{N_{f}}m_{a}\right)\label{eq:N4_Ugly_Hyperbolic}
\end{equation}
and we identify

\[
Z_{\text{background FI}}\left(\{m_{a}\};\eta\right)=c\left(4\eta\sum_{a=1}^{N_{f}}m_{a}\right)
\]
the decoupled sector represented by $Z_{\text{hyper}}^{\mathcal{N}=4}\left(\eta\right)$
is associated with the free twisted hypermultiplet which is the dual
of the dimension $1/2$ monopole. 

In trying to identify the duality implied by the identity for ``bad''
theories we run into an ambiguity in the interpretation of the factor
\[
\Gamma_{h}\left((m+1)\omega-\frac{1}{2}\sum_{a=1}^{n+m}(\mu_{a}+\nu_{a})\pm\frac{1}{2}\lambda\right)=\Gamma_{h}\left(\left(\frac{N_{f}}{2}-N_{c}+1\right)\omega\pm\eta\right)
\]
An interpretation of this factor as arising from a single (twisted)
hypermultiplet (or two chiral multiplets as in \cite{Willett:2011gp})
is inconsistent. In order to bypass this difficulty, we begin with
the duality for the ``ugly'' theory and proceed by ``integrating
out'' matter multiplets on both sides, leading to a duality for ``bad''
theories. We will shortly clarify what integrating out means in the
context of the matrix model.

\subsection{Integrating out flavors}

We would like to define a procedure for integrating out flavors in
the $S_{b}^{3}$ partition function, starting from the ``ugly''
theory with $N_{f}=2N_{c}-1$. In principle, this can be done by taking
some of the mass parameters to infinity. Naive application of this
type of limit leads to an inconsistency of the type noted in \ref{sub:Flowing-from-ugly}.
Instead, we will try to mimic the physically acceptable picture of
partial Higgsing in the matrix model. This can be done by allowing
some of the integration variables to transform along with the mass
parameters \cite{Aharony:toappear}. The number of such variables
is a priori arbitrary. We will use the asymptotic behavior of the
matrix integral, as we take the mass parameter to infinity, as a guide
for identifying the correct vacuum%
\footnote{We would like to thank Brian Willett for introducing us to this technique.%
}. A related ``degeneration'' procedure is described in \cite{VanDeBult2008}
(see also \cite{Benini:2011mf}).

As shown in \cite{VanDeBult2008} (5.2.6), the hyperbolic gamma function
has the following asymptotic behavior

\[
\Gamma_{h}(x|\omega_{1},\omega_{2})\approx\exp\left(\pm2\pi i\left(\frac{\left(x-\omega\right)^{2}}{4\omega_{1}\omega_{2}}-\frac{\omega_{1}^{2}+\omega_{2}^{2}}{48\omega_{1}\omega_{2}}\right)\right)=\left(\zeta^{-1}c\left(\left(x-\omega\right)^{2}\right)\right)^{\pm1}\qquad x\rightarrow\pm\infty
\]
\[
\zeta=\exp\left(\pi i\left(\frac{\omega_{1}^{2}+\omega_{2}^{2}}{24\omega_{1}\omega_{2}}\right)\right)
\]
We take a single mass parameter $m_{1}=\xi$ to be very large and
allow $\alpha$ of the integration variables to have a shift which
cancels this particular parameter. The resulting integral (already
at the $\mathcal{N}=4$ values) is

\begin{align}
I_{n,(2,2)}^{m}(\mu;\nu;\lambda)= & \sum_{\alpha=1}^{n}\left(\begin{array}{c}
n\\
\alpha
\end{array}\right)\frac{1}{\sqrt{-\omega_{1}\omega_{2}}^{n}n!}\int_{C^{n}}\prod_{1\leq j<k\leq\alpha}\frac{1}{\Gamma_{h}(\pm(x_{j}-x_{k}))}\prod_{\alpha+1\leq j<k\leq n}\frac{1}{\Gamma_{h}(\pm(x_{j}-x_{k}))}\notag\nonumber \\
 & \prod_{1\leq j\le\alpha,\alpha+1\le k\le n}\frac{1}{\Gamma_{h}(\pm(x_{j}-\xi-x_{k}))}\notag\nonumber \\
 & \prod_{j=\alpha+1}^{n}\left(e^{\frac{-2\pi i\eta x_{j}}{\omega_{1}\omega_{2}}}\Gamma_{h}(\frac{\omega}{2}-\xi-x_{j})\Gamma_{h}(\frac{\omega}{2}+\xi+x_{j})\prod_{a=2}^{n+m}\Gamma_{h}(\frac{\omega}{2}-m_{a}-x_{j})\Gamma_{h}(\frac{\omega}{2}+m_{a}+x_{j})\right)dx_{j}\notag\nonumber \\
 & \prod_{j=1}^{\alpha}\left(e^{\frac{-2\pi i\eta(x_{j}-\xi)}{\omega_{1}\omega_{2}}}\Gamma_{h}(\frac{\omega}{2}-x_{j})\Gamma_{h}(\frac{\omega}{2}+x_{j})\prod_{a=2}^{n+m}\Gamma_{h}(\frac{\omega}{2}-m_{a}-x_{j}+\xi)\Gamma_{h}(\frac{\omega}{2}+m_{a}+x_{j}-\xi)\right)dx_{j}
\end{align}
where the first factor arises from the choice of $\alpha$ out of
$n$ variables. Using the asymptotic form for the hyperbolic gamma
function

\begin{align}
I_{n,(2,2)}^{m}(\mu;\nu;\lambda)\approx & \sum_{\alpha=1}^{n}\frac{\left(\begin{array}{c}
n\\
\alpha
\end{array}\right)}{\sqrt{-\omega_{1}\omega_{2}}^{n}n!}\int_{C^{n}}\Bigg\{\prod_{1\leq j<k\leq\alpha}\frac{1}{\Gamma_{h}(\pm(x_{j}-x_{k}))}\prod_{\alpha+1\leq j<k\leq n}\frac{1}{\Gamma_{h}(\pm(x_{j}-x_{k}))}\nonumber \\
 & \prod_{1\leq j\le\alpha,\alpha+1\le k\le n}c\left(\left(x_{j}-\xi-x_{k}-\omega\right)^{2}-\left(x_{j}-\xi-x_{k}+\omega\right)^{2}\right)\nonumber \\
 & \prod_{j=1}^{\alpha}\Bigg[e^{\frac{-2\pi i\eta(x_{j}-\xi)}{\omega_{1}\omega_{2}}}\prod_{a=2}^{n+m}c\left(\left(\frac{\omega}{2}-m_{a}-x_{j}+\xi-\omega\right)^{2}-\left(\frac{\omega}{2}+m_{a}+x_{j}-\xi-\omega\right)^{2}\right)\times\nonumber \\
 & \Gamma_{h}(\frac{\omega}{2}-x_{j})\Gamma_{h}(\frac{\omega}{2}+x_{j})dx_{j}\Bigg]\nonumber \\
 & \prod_{j=\alpha+1}^{n}\Bigg[\left(e^{\frac{-2\pi i\eta x_{j}}{\omega_{1}\omega_{2}}}\prod_{a=2}^{n+m}\Gamma_{h}(\frac{\omega}{2}-m_{a}-x_{j})\Gamma_{h}(\frac{\omega}{2}+m_{a}+x_{j})\right)\times\nonumber \\
 & c\left(\left(\frac{\omega}{2}+\xi+x_{j}-\omega\right)^{2}-\left(\frac{\omega}{2}-\xi-x_{j}-\omega\right)^{2}\right)dx_{j}\Bigg]\Bigg\}
\end{align}
The various factors simplify (for $\alpha>0$) as

\begin{align}
\prod_{1\leq j<\alpha,\alpha+1\le k\le n}c\left(\left(x_{j}-\xi-x_{k}-\omega\right)^{2}-\left(x_{j}-\xi-x_{k}+\omega\right)^{2}\right)\nonumber \\
=\prod_{1\leq j<\alpha,\alpha+1\le k\le n}c\left(4\omega\left(x_{k}-x_{j}+\xi\right)\right)=\nonumber \\
=c\left(4\omega\xi\alpha\left(n-\alpha\right)\right)\prod_{1\leq j\le\alpha}c\left(-4\omega x_{j}(n-\alpha)\right)\prod_{\alpha+1\le k\le n}c\left(4\omega x_{k}\alpha\right)
\end{align}

\begin{align}
\prod_{j=\alpha+1}^{n}c\left(\left(\frac{\omega}{2}+\xi+x_{j}-\omega\right)^{2}-\left(\frac{\omega}{2}-\xi-x_{j}-\omega\right)^{2}\right) & =\prod_{j=\alpha+1}^{n}c\left(-2\omega\left(x_{j}+\xi\right)\right)=c\left(-2\omega\left(n-\alpha\right)\xi\right)\times\nonumber \\
 & \times\prod_{j=\alpha+1}^{n}c\left(-2\omega x_{j}\right)
\end{align}

\begin{align}
\prod_{j=1}^{\alpha}\left(e^{\frac{-2\pi i\eta(x_{j}-\xi)}{\omega_{1}\omega_{2}}}\prod_{a=2}^{n+m}c\left(\left(\frac{\omega}{2}-m_{a}-x_{j}+\xi-\omega\right)^{2}-\left(\frac{\omega}{2}+m_{a}+x_{j}-\xi-\omega\right)^{2}\right)\right)\nonumber \\
=\prod_{j=1}^{\alpha}\left(e^{\frac{-2\pi i\eta(x_{j}-\xi)}{\omega_{1}\omega_{2}}}\prod_{a=2}^{n+m}c\left(2\omega\left(m_{a}+x_{j}-\xi\right)\right)\right)\nonumber \\
=c\left(4\alpha\eta\xi\right)c\left(-2\omega\xi\left(n+m-1\right)\alpha+2\omega\alpha\sum_{a=2}^{n+m}m_{a}\right)\prod_{j=1}^{\alpha}c\left(x_{j}\left(-4\eta+\left(n+m-1\right)\omega\right)\right)
\end{align}
The result is then the following sum over $\alpha$ sectors

\begin{align}
I_{n,(2,2)}^{m}\left(\xi,\frac{\omega}{2}-m_{a};\xi,\frac{\omega}{2}+m_{a};-2\eta\right) & \approx I_{n,(2,2)}^{m-1}\left(\frac{\omega}{2}-m_{a};\frac{\omega}{2}+m_{a};-2\eta+\omega\right)c\left(-2\omega\xi n\right)+\nonumber \\
 & \sum_{\alpha=1}^{n}\Bigg\{ I_{\alpha,(2,2)}^{1-\alpha}\left(\frac{\omega}{2};\frac{\omega}{2};-2\eta+(-n+m-1+2\alpha)\omega\right)\times\nonumber \\
 & \qquad I_{n-\alpha,(2,2)}^{m+\alpha-1}\left(\frac{\omega}{2}-m_{a};\frac{\omega}{2}+m_{a};-2\eta+\omega\right)\times\nonumber \\
 & \qquad\times c\left(2\omega\xi\left(n\left(\alpha-1\right)-\alpha\left(m+2\alpha-2\right)\right)+4\alpha\eta\xi+2\omega\alpha\sum_{a=2}^{n+m}m_{a}\right)\Bigg\}\label{eq:alpha_sectors}
\end{align}
Note that each element in the sum includes the partition function
of the original theory Higgsed to $U(\alpha)\times U(n-\alpha)$ as
required on physical grounds.

\subsection{Matching}

$ $We now attempt to match vacua between the ``ugly'' theory and
its dual with one flavor integrated out on both sides. That is, we
will solve for the Higgsing of the dual theory in terms of the original
one.

We are seeking a solution where the original gauge group is unbroken,
hence $\alpha=0$. The scaling with $\xi$ of the $\alpha=0$ summand
in the ``ugly'' theory ($N_{f}=2N_{c}-1$) is
\[
c\left(-2\omega\xi N_{c}\right)
\]
and an $\tilde{\alpha}$ sector in the dual ($\tilde{N}_{c}=N_{c}-1,\tilde{N}_{f}=2N_{c}-1$)
\[
c\left(-2\omega\xi\left(N_{c}-\tilde{\alpha}-1+2\tilde{\alpha}^{2}\right)\right)
\]
The only integral solution for which the scaling matches is $\tilde{\alpha}=1$.
Hence the gauge group of the dual is spontaneously broken in the right
vacuum to $U(N_{c}-2)\times U(1)$. The second factor has a single
massless charged hypermultiplet. We must also check to see that the
$\eta$ dependent $\xi$ scaling of the two theories matches. The
``ugly'' theory, having $\alpha=0$ does not scale. For the dual,
the above derivation gives a factor of (recall that the parameter
$\eta$ appearing in the integral for the dual is $-\eta$ of the
original integral) 
\[
c\left(-4\tilde{\alpha}\eta\xi\right)
\]
There is one additional contribution coming from the duality relationship
\ref{eq:duality_relation} 
\[
c\left(4\eta\sum_{a=1}^{N_{f}}m_{a}\right)=c\left(4\eta\xi\right)c\left(4\eta\sum_{a=2}^{N_{f}}m_{a}\right)
\]
which exactly cancels the first factor for $\tilde{\alpha}=1$.

At this point we appeal to abelian mirror symmetry of $\mathcal{N}=4$
theories which relates a $U(1)$ theory with a single massless charged
flavor to a free twisted hypermultiplet. The real mass parameter of
the twisted hypermultiplet is mapped to the FI term of the gauge theory.
In terms of the hyperbolic gamma function integrals, this is nothing
but the $n=1,m=0$ version of \ref{eq:duality_relation} which states
\begin{equation}
I_{1,(2,2)}^{0}\left(\frac{\omega}{2}-m_{a};\frac{\omega}{2}+m_{a};-2\eta\right)=\Gamma_{h}\left(\frac{\omega}{2}\pm\eta\right)c\left(4\eta m\right)\label{eq:N4_Ugly_Hyperbolic-1}
\end{equation}
equivalently 
\begin{equation}
Z_{1,1}^{\mathcal{N}=4}\left(m;\eta\right)=Z_{\text{hyper}}^{\mathcal{N}=4}\left(\eta\right)Z_{\text{background FI}}\left(m;\eta\right)\label{eq:abelian_mirror}
\end{equation}
and we need to consider the situation where $m=0$.

Combining \ref{eq:duality_relation}, \ref{eq:alpha_sectors} and
\ref{eq:abelian_mirror}, and taking $\alpha=0,\tilde{\alpha}=1$,
we conclude that 
\begin{align}
Z_{N_{c},2N_{c}-2}^{\mathcal{N}=4}\left(\{m_{a}\};\eta+\omega/2\right) & =Z_{N_{c}-2,2N_{c}-2}^{\mathcal{N}=4}\left(\{m_{a}\};-\eta-\omega/2\right)Z_{\text{hyper}}^{\mathcal{N}=4}\left(-\eta\right)\times\nonumber \\
 & \times Z_{\text{hyper}}^{\mathcal{N}=4}\left(\eta\right)Z_{\text{background FI}}\left(\{m_{a}\};\eta+\frac{\omega}{2}\right)
\end{align}
or, defining $\eta_{r}=\eta+\omega/2$
\begin{align}
Z_{N_{c},2N_{c}-2}^{\mathcal{N}=4}\left(\{m_{a}\};\eta_{r}\right) & =Z_{N_{c}-2,2N_{c}-2}^{\mathcal{N}=4}\left(\{m_{a}\};-\eta_{r}\right)Z_{\text{hyper}}^{\mathcal{N}=4}\left(-\eta_{r}+\omega/2\right)\times\label{eq:Next_to_ugly-2}\\
 & \times Z_{\text{hyper}}^{\mathcal{N}=4}\left(\eta_{r}-\omega/2\right)Z_{\text{background FI}}\left(\{m_{a}\};\eta_{r}\right)
\end{align}
which has the desired form. 

What happens if we continue to integrate out flavors? Assuming that
the pattern holds after integrating out $1<k<N_{c}-1$ flavors the
scaling of the $U(N_{c})$ theory with $2N_{c}-k$ flavors is 
\[
c\left(-2\omega\xi N_{c}\right)
\]
and that of the $U(N_{c}-k)$ theory with $2N_{c}-k$ flavors 
\[
c\left(2\omega\xi\left(\left(N_{c}-k\right)\left(\tilde{\alpha}-1\right)-\tilde{\alpha}\left(N_{c}+2\tilde{\alpha}-2\right)\right)\right)
\]
so that again $\tilde{\alpha}=1$ and the pattern continues until
$k=N_{c}-1$, at which point the dual gauge group is trivial and the
process terminates at the $N_{f}=N_{c}$ theory. Combining these results
we conclude that for $N_{c}\le N_{f}<2N_{c}$
\begin{equation}
Z_{N_{c},N_{f}}^{\mathcal{N}=4}\left(\{m_{a}\};\eta\right)=Z_{N_{f}-N_{c},N_{f}}^{\mathcal{N}=4}\left(\{m_{a}\};-\eta\right)Z_{\text{background FI}}\left(\{m_{a}\};\eta\right)Z_{\text{hyper}}^{\mathcal{N}=4}\left(\eta_{u}\right)\prod_{j=1}^{2N_{c}-N_{f}-1}Z_{\text{hyper}}^{\mathcal{N}=4}\left(\eta_{j}\right)\label{eq:Next_to_ugly-2-1}
\end{equation}
At each step in the process of integrating out flavors, the FI parameter
$\eta$ will shift by $\omega/2$ (there are $2N_{c}-N_{f}-j$ such
steps). Together with the number of flavors at each step, this sets
the values of the hypermultiplet deformation parameters to
\[
\eta_{u}=\eta-\left(2N_{c}-N_{f}-1\right)\frac{\omega}{2},\qquad\eta_{j}=-\eta+\left(2N_{c}-N_{f}+1-2j\right)\frac{\omega}{2}
\]

\subsection{Interpretation}

We now reconsider the ambiguous factor in the duality relation \ref{eq:duality_relation}
\[
\Gamma_{h}\left(\left(\frac{N_{f}}{2}-N_{c}+1\right)\omega\pm\eta\right)
\]
First, we would like to show that indeed
\begin{equation}
\Gamma_{h}\left(\left(\frac{N_{f}}{2}-N_{c}+1\right)\omega\pm\eta\right)=\prod_{i=1}^{2N_{c}-N_{f}}\Gamma_{h}\left(\frac{\omega}{2}\pm\eta_{j}\right)\label{eq:decoupled_sector_identity}
\end{equation}
where we have combined the product. Recall that
\[
Z_{\text{hyper}}^{\mathcal{N}=4}\left(\eta\right)=\Gamma_{h}\left(\frac{\omega}{2}\pm\eta\right)
\]
It is easily checked that the product on the rhs of \ref{eq:decoupled_sector_identity}
is mostly telescopic with respect to the identity
\begin{equation}
\Gamma_{h}\left(z\right)\Gamma_{h}\left(2\omega-z\right)=1\label{eq:hyperbolic_gamma_inverse}
\end{equation}
and that the remainder, after cancelations, is exactly the lhs. We
interpret this to mean that the ambiguous factor represents the contribution
of $2N_{c}-N_{f}$ hypermultiplets. The relation \ref{eq:duality_relation}
then supports our duality proposal for all the relevant values of
$N_{c}$ and $N_{f}$. 

The part of the deformation parameters $\eta_{j}$ which depends on
$\omega$ represents a non-canonical R-charge for the free hypermultiplets.
Moreover, the R-charge is different for the two chiral fields in each
hypermultiplet. Assuming the duality holds, we could set these R-charges
to $1/2$ and recover the partition function for the original theory,
but with the correct R-charge assignment for all the fields. For instance,
we could compute the entanglement entropy of the ``bad'' theory.
There is no dual deformation in the integral expression for the original
theory which would allow us to do this.

\section{Discussion and conclusions}

We have given some evidence in support of a proposal for a Seiberg-like
duality of 3d $\mathcal{N}=4$ gauge theories with $N_{c}\le N_{f}<2N_{c}$.
The proposed dual has the advantage that some of the accidental symmetries
present in the IR, those associated with the free hypermultiplets,
are manifest. One could, in principle, search for these currents in
the original theory by considering monopole operators. Such symmetry
enhancement by monopole operators has been shown to exist in several
examples \cite{Bashkirov:2010hj,Bashkirov:2010kz,Gaiotto:2008ak}.
The analysis in ``bad'' theories is complicated by the fact that
the R-charges for monopole operators are corrected by (we assume)
the very accidental symmetries we are searching for. It may still
be possible to describe a self consistent scenario for the monopole
operator spectrum without the need to identify an explicit dual. 

The existence of a dual with enough visible symmetry allows us to
compute expectation values for physical observable of a ``bad''
theory. This includes the entanglement entropy (i.e. the $S^{3}$
partition function \cite{Casini:2011kv}) and the expectation values
of various supersymmetric Wilson loops \cite{Kapustin:2009kz} or
defect operators \cite{Kapustin:2012iw,Drukker:2012sr}. The dual
operators for the latter two must still be identified as was done
for Seiberg-like Chern-Simons duals in \cite{Kapustin:2013hpk}. 

We have argued that at a generic point on the mixed Higgs-Coulomb
branch, strong dynamics require the introduction of additional hypermultiplets
in order to describe the low energy theory, along the same lines as
in $\mathcal{N}=2$ \cite{Aharony:1997bx}. An alternative approach
to calculating the size of a possible decoupled free sector would
be to examine the corrections to the metric on the Coulomb branch.
The corrected metric may be quite complicated, but the splitting of
the moduli space into a flat space parametrizing the free sector and
the Coulomb branch of the dual theory should be visible even far out
along this space where an analysis based on instantons (and one loop
diagrams) along the lines of \cite{Seiberg:1996nz} is applicable.
By doing this, one may be able to see the spontaneous breaking of
the Weyl group of the original gauge group.

\acknowledgments

I would like to thank Ofer Aharony, Brian Willett, Anton Kapustin
and Davide Gaiotto for very useful discussions and comments on the
draft. I would especially like to thank the authors of \cite{Aharony:toappear}
for making a part of their work available before publication. This
research is supported by NSF grant PHY-0756966.

\bibliographystyle{JHEP}
\bibliography{General_bibliography}

\end{document}